\documentclass[%
 reprint,
 superscriptaddress,
 amsmath,amssymb,
 aps,
 noeprint,
]{revtex4-2}
\usepackage{hyperref}
\hypersetup{
colorlinks=true,
linkcolor=blue,
urlcolor=black,
citecolor=blue,
anchorcolor=blue
}
\usepackage{graphicx}
\usepackage{svg}
\usepackage{float}
\usepackage{physics}
\usepackage{dcolumn}
\usepackage{bm}
\allowdisplaybreaks 

\begin{document}

\preprint{AAPM/123-QED}

\title{A high-fidelity two-qubit gate for multimode superconducting P-mon qubits}

\author{F.~Pfeiffer}
\email{frederik.pfeiffer@wmi.badw.de}
\affiliation{Technical University of Munich, TUM School of Natural Sciences, Department of Physics, Garching 85748, Germany}
\affiliation{Walther-Mei\ss ner-Institut, Bayerische Akademie der Wissenschaften, Garching 85748, Germany}

\author{F.~A.~Roy}
\affiliation{Walther-Mei\ss ner-Institut, Bayerische Akademie der Wissenschaften, Garching 85748, Germany}
\affiliation{Theoretical Physics, Saarland University, 66123 Saarbrücken, Germany}

\author{N.~J.~Glaser}
\affiliation{Technical University of Munich, TUM School of Natural Sciences, Department of Physics, Garching 85748, Germany}
\affiliation{Walther-Mei\ss ner-Institut, Bayerische Akademie der Wissenschaften, Garching 85748, Germany}

\author{J.~Feigl}
\affiliation{Technical University of Munich, TUM School of Natural Sciences, Department of Physics, Garching 85748, Germany}
\affiliation{Walther-Mei\ss ner-Institut, Bayerische Akademie der Wissenschaften, Garching 85748, Germany}

\author{L.~Koch}
\affiliation{Technical University of Munich, TUM School of Natural Sciences, Department of Physics, Garching 85748, Germany}
\affiliation{Walther-Mei\ss ner-Institut, Bayerische Akademie der Wissenschaften, Garching 85748, Germany}

\author{K.~Kiener}
\affiliation{Technical University of Munich, TUM School of Natural Sciences, Department of Physics, Garching 85748, Germany}
\affiliation{Walther-Mei\ss ner-Institut, Bayerische Akademie der Wissenschaften, Garching 85748, Germany}

\author{G.~Krylov}
\affiliation{Technical University of Munich, TUM School of Natural Sciences, Department of Physics, Garching 85748, Germany}
\affiliation{Walther-Mei\ss ner-Institut, Bayerische Akademie der Wissenschaften, Garching 85748, Germany}

\author{J.~Schirk}
\affiliation{Technical University of Munich, TUM School of Natural Sciences, Department of Physics, Garching 85748, Germany}
\affiliation{Walther-Mei\ss ner-Institut, Bayerische Akademie der Wissenschaften, Garching 85748, Germany}

\author{\\C. M. F. ~Schneider}
\affiliation{Technical University of Munich, TUM School of Natural Sciences, Department of Physics, Garching 85748, Germany}
\affiliation{Walther-Mei\ss ner-Institut, Bayerische Akademie der Wissenschaften, Garching 85748, Germany}

\author{L.~S\"odergren}
\affiliation{Technical University of Munich, TUM School of Natural Sciences, Department of Physics, Garching 85748, Germany}
\affiliation{Walther-Mei\ss ner-Institut, Bayerische Akademie der Wissenschaften, Garching 85748, Germany}

\author{F.~Wallner}
\affiliation{Technical University of Munich, TUM School of Natural Sciences, Department of Physics, Garching 85748, Germany}
\affiliation{Walther-Mei\ss ner-Institut, Bayerische Akademie der Wissenschaften, Garching 85748, Germany}

\author{M.~Werninghaus}
\affiliation{Technical University of Munich, TUM School of Natural Sciences, Department of Physics, Garching 85748, Germany}
\affiliation{Walther-Mei\ss ner-Institut, Bayerische Akademie der Wissenschaften, Garching 85748, Germany}

\author{C. A. Riofrío}
\affiliation{BMW Group, Munich, Germany}

\author{S.~Filipp}
\email{stefan.filipp@wmi.badw.de}
\affiliation{Technical University of Munich, TUM School of Natural Sciences, Department of Physics, Garching 85748, Germany}
\affiliation{Walther-Mei\ss ner-Institut, Bayerische Akademie der Wissenschaften, Garching 85748, Germany}

\date{\today}

\begin{abstract}
To scale superconducting quantum processors, it is essential to achieve long coherence times while engineering interactions that do not introduce additional decoherence channels. In superconducting qubit systems, this can be realized using multimode circuits that feature a protected qubit mode alongside a distinct mediator mode. Building on this concept, our recently developed P-mon qubit provides intrinsic protection against decoherence from the readout environment. We extend this approach to controlled two-qubit interactions, by exploiting the mediator modes of P-mons for on-demand coupling. Because direct interactions between the qubit modes are strongly suppressed, unwanted $ZZ$-type interactions are significantly reduced to below $3.6(5)~\text{kHz}$ in the idle state. 
When tuning the coupled mediator modes on resonance, the cross-Kerr interaction between the qubit and the hybridized mediator modes leads to a qubit-state dependent frequency shift. By selectively addressing these transitions, we implement a $180~\text{ns}$ long CZ gate and determine a fidelity of $99.62(4)~\text{\%}$. These results represent a significant step toward a scalable superconducting architecture that maintains high performance at scale.
\end{abstract}

\maketitle

\section{Introduction}
Superconducting qubits have emerged as a leading platform for realizing fault-tolerant quantum computing \cite{Krinner2022, Acharya2023, Acharya2025}. As fabrication techniques continue to mature \cite{Place2021, Tuokkola2025, Bland2025, Bruckmoser2026}, the central challenge is to develop scalable processor architectures that preserve the high performance of individual qubits when integrated into large-scale arrays. A key issue is the emergence of additional error channels. Among these, residual qubit-qubit ZZ-type interactions are particularly problematic, as they induce correlated--always-on--coherent errors \cite{Gambetta2012, Mundada2019, Krinner2020}.

In the standard transmon architecture \cite{Koch2007}, interactions are typically realized through capacitive couplings, resulting in linear exchange terms. Parasitic $ZZ$ interactions then arise from interaction with higher-energy levels of the transmon \cite{DiCarlo2009, Barends2014}. While such unwanted interactions can be mitigated using frequency-tunable qubits \cite{DiCarlo2009}, this approach increases susceptibility to flux-noise and does not fully eliminate $ZZ$ coupling. Tunable couplers to modulate the linear interaction \cite{Chen2014, McKay2016} can partially solve this issue. Although these allow for sufficient $ZZ$ suppression and high-fidelity gates, they typically require operating transmon-type qubits in the so-called ``straddling regime'' \cite{Koch2007}, which leads to frequency crowding and increased susceptibility to microwave crosstalk. One strategy to relax these constraints is to use more advanced coupler architectures \cite{Hayato2022, Li2024, An2025}. 

Rather than increasing hardware complexity through additional coupling elements, an alternative strategy is to internalize the coupling mechanism by re-engineering the qubit itself. Here, we follow such an approach and extend the transmon to a multimode circuit, which, in addition to a qubit mode, includes a mediator mode. This mode serves as an indirect link to readout circuits and other qubits. Nonlinear cross-Kerr interactions between the modes occur in these circuits naturally through shared Josephson junction nonlinearities \cite{Gambetta2011, Hoffman2011, Srinivasan2011, Roy2017, Roy2018}, a resource not accessible via purely capacitive couplings. While such circuits have primarily been used to implement nonlinear interactions with readout resonators to enhance protection and readout performance \cite{Zhang2017, Dassonneville2020, Dassonneville2023, Hazra2025, Kline2025, Salunkhe2025, Mori2025, Mori2025_2, Hazra2026}, they also enable an interaction structure that avoids linear interactions between computational modes entirely. 

This concept has been demonstrated for the tunable coupling qubit (TCQ) \cite{Finck2021}, which achieved a $740~\text{ns}$ resonator-induced phase (RIP) gate by utilizing a second-order dispersive interaction, a technique previously developed for transmons \cite{Cross2015, Paik2016}. However, the TCQ architecture faces inherent trade-offs in noise resilience. The qubit mode is protected from Purcell decay, but cannot simultaneously be protected from photon-induced dephasing. Moreover, introducing magnetic-flux control of the mediator mode renders the qubit mode susceptible to flux-noise. 

To circumvent these limitations, we have introduced the multimode P-mon circuit \cite{Pfeiffer2024}, which features a flux-tunable mediator mode and a flux-noise insensitive qubit mode. In that work, we used the mediator mode to mediate the interaction of the qubit mode to its readout resonator, showing that it can be protected simultaneously against both Purcell decay and photon-induced dephasing. Here, we extend these protection mechanisms to qubit-qubit couplings and show that $ZZ$-type crosstalk can be efficiently suppressed. By linking the mediator modes of two P-mons via a linear exchange coupling--while the qubit modes remain decoupled--we implement a two-qubit gate scheme that exploits non-perturbative cross-Kerr interactions directly rather than second-order dispersive interactions. When the mediator modes are tuned into resonance, the internal cross-Kerr interactions induce a qubit-state-dependent shift in the hybridized mediator frequencies. Consequently, while the qubits remain effectively isolated when the mediators are in their ground states, driving a state-specific transition generates a cyclic evolution [see Fig.~\ref{fig:circuit}(a)] resulting in a state-dependent phase accumulation and a controlled-phase (CPHASE) gate.

\section{Coupled P-mon Architecture}
\begin{figure}[H]
    \centering
    \includegraphics{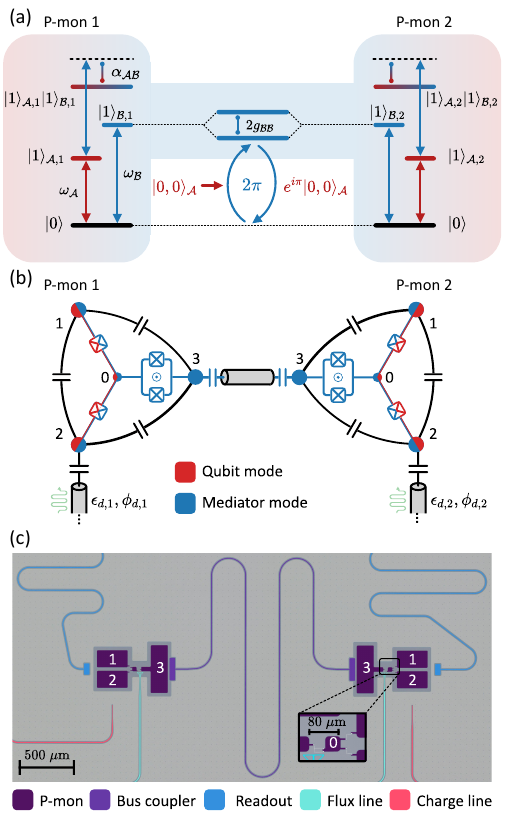}      
 \caption{(a) Energy level diagram of two P-mon qubits coupled via mediator modes. $\omega_\mathcal{A}$ and $\omega_\mathcal{B}$ are the transition frequencies for single excitation states $|1\rangle_{\mathcal{A},1} \equiv |1, 0\rangle_\mathcal{A} \otimes |0, 0\rangle_\mathcal{B}$ and $|1\rangle_{\mathcal{B},1} \equiv |0, 0\rangle_\mathcal{A} \otimes |1, 0\rangle_\mathcal{B}$ of modes $\mathcal{A}$ and $\mathcal{B}$ of P-mon 1 (and similarly $|1\rangle_{\mathcal{A},2}$ and $|1\rangle_{\mathcal{B},2}$ denote the excitation states of P-mon 2). The cross-Kerr interaction $\alpha_\mathcal{AB}$ shifts the energy of the joint excitation state $|1\rangle_{\mathcal{A},1} |1\rangle_{\mathcal{B},1} \equiv |1, 0\rangle_\mathcal{A} \otimes |1, 0\rangle_\mathcal{B}$. The spectrum of the coupled mediators (coupling strength $g_\mathcal{BB}$) is sketched for the qubit modes ground state $|0,0\rangle_\mathcal{A}$; a selective cyclic evolution on this state yields a geometric phase of $\pi$. (b) Circuit diagram of two coupled P-mon circuits. Each four-island P-mon is connected via two Josephson junctions and a SQUID. Capacitances to the center island $C_{0i}$ and ground $C_{gj}$ are omitted. A charge line couples to island 2 of each P-mon, while a coplanar waveguide resonator capacitively couples to both via island 3. (c) False colored microscope image of the device. The numbers label the circuit islands corresponding to the labels in (b).}
\label{fig:circuit}
\end{figure}

Each P-mon is a four node circuit, as illustrated in the circuit schematic in Fig.~\ref{fig:circuit}(b). Charge oscillations between islands 1 and 2 via the two Josephson junctions form the fixed frequency qubit mode $\mathcal{A}$. The mediator mode $\mathcal{B}$ between islands 1, 2 and 3 also involves the SQUID loop, making its frequency flux-tunable. While a third mode $\mathcal{C}$ is inherently present in the circuit, it can be decoupled by shifting it to high frequencies through a choice of circuit parameters \cite{Pfeiffer2024}. The nonlinearity of the Josephson junctions gives rise to a cross-Kerr interaction $\alpha_\mathcal{AB}$, which provides
a coupling between the qubit and the mediator mode within a single P-mon; as a result, the mediator's resonance frequency becomes dependent on the qubit's excitation state. Since island 3 participates exclusively in mode $\mathcal{B}$, it can be used to selectively couple to the mediator mode without introducing coupling to the qubit mode $\mathcal{A}$. Consequently, capacitively coupling two P-mon circuits via their respective island 3 induces a linear coupling between the mediators while preventing direct qubit-qubit interaction. Here, we utilize an intermediate bus mode to virtually mediate this interaction \cite{Majer2007, Fillip2011_2} instead of a direct capacitive coupling to simplify physical routing.

For a quantitative understanding, we first consider the effective Hamiltonian of a single P-mon circuit,
\begin{align}
    \hat{H}_{P} =& \sum_m \omega_m \hat{a}_m^\dagger \hat{a}_m + \frac{\alpha_m}{2} \hat{a}_m^\dagger \hat{a}_m^\dagger \hat{a}_m \hat{a}_m \nonumber \\  &+ \alpha_{\mathcal{AB}} \hat{a}_\mathcal{A}^\dagger \hat{a}_\mathcal{A} \hat{a}_\mathcal{B}^\dagger \hat{a}_\mathcal{B}, \label{eq:pmon_effective_hamiltonian_no_asymmetry}
\end{align}
where $\hat{a}_m$ are the bosonic annihilation operators for the qubit and mediator modes ($m=\mathcal{A}, \mathcal{B}$), $\omega_m$ their respective frequencies, $\alpha_m$ their self-Kerr and $\alpha_{\mathcal{AB}}$ their cross-Kerr interaction. Note that this expression neglects the third high-frequency mode $\mathcal{C}$ present in the circuit \cite{Pfeiffer2024}. 

When two such circuits are coupled via an intermediate bus resonator of frequency $\omega_r$, the total system Hamiltonian becomes
\begin{align}
    \hat{H} = \sum_\mu \hat{H}_{P\mu} + \omega_r \hat{a}_r^\dagger \hat{a}_r  + \sum_\mu g_{\mathcal{B}r, \mu} \left( \hat{a}_{\mathcal{B}\mu}^\dagger \hat{a}_r + \text{h.c.} \right), \label{eq:pmons_bus_hamiltonian}
\end{align}
\noindent where $g_{\mathcal{B}r, \mu}$ denotes the coupling strength between the $\mu$-th mediator and the bus. A detailed derivation based on the underlying circuit model is provided in Appendix~\ref{app:effective_hamiltonian}.
We operate in the dispersive limit, where the bus resonator is far detuned from the mediator modes ($|\omega_{\mathcal{B},i} - \omega_r| \gg g_{\mathcal{B}r,i}$), and eliminate the bus resonator via a Schrieffer-Wolff transformation \cite{Majer2007, Blais2021}.
This transformation reduces the system to the composite Hilbert space of the qubit and mediator modes, $\mathcal{H} = \mathcal{H}_\mathcal{A} \otimes \mathcal{H}_\mathcal{B}$, spanned by the basis $\mathcal{S} = \{ |i,j\rangle_\mathcal{A} \otimes |k,l\rangle_\mathcal{B} : i,j,k,l \in \mathbb{N} \}$.

By further truncating the qubit and mediator modes to their single-excitation subspaces ($i,j,k,l \in \{0, 1\}$), we find the effective interaction Hamiltonian:
\begin{align}
    \hat{H} &= \sum_\mu \omega_{\mathcal{A}\mu} \hat{\sigma}_{\mathcal{A}\mu}^\dagger \hat{\sigma}_{\mathcal{A}\mu} + (\omega_{\mathcal{B}\mu} + \alpha_{\mathcal{A}\mathcal{B}\mu} \hat{\sigma}_{\mathcal{A}\mu}^\dagger \hat{\sigma}_{\mathcal{A}\mu}) \hat{\sigma}_{\mathcal{B}\mu}^\dagger \hat{\sigma}_{\mathcal{B}\mu} \nonumber \\
    &+ g_{\mathcal{BB}} ( \hat{\sigma}_{\mathcal{B}1}^\dagger \hat{\sigma}_{\mathcal{B}2} + \text{h.c.}), \label{eq:H_AB_TLS}
\end{align}
where $\sigma_{\mathcal{A}\mu}$ and $\sigma_{\mathcal{B}\mu}$ are the Pauli operators of the truncated subspace, and $g_\mathcal{BB} = g_{\mathcal{B}r,1}g_{\mathcal{B}r,2} (1/ \Delta_1 + 1/ \Delta_2)/2$ is the bus-mediated effective mediator-mediator coupling with detunings $\Delta_{1/2} = \omega_{\mathcal{B},1/2} - \omega_r$. 

Eq.~\eqref{eq:H_AB_TLS} highlights the two key ideas of the coupling scheme. First, the computational qubit modes $\mathcal{A}_{1/2}$ (first term) do not interact via a linear interaction; they interact only with their local mediators $\mathcal{B}_{1/2}$ via the second, cross-Kerr term. Consequently, if the mediators remain in their ground state ($ \langle \sigma_{\mathcal{B}, i}^\dagger \sigma_{\mathcal{B}, i} \rangle = 0$), the qubits are completely decoupled. Second, the cross-Kerr term $\alpha_{\mathcal{AB}, i}$ acts as a qubit-state-dependent frequency shift for the mediators and forms the basis for the mediator-activated CPHASE gate discussed in Sec.~\ref{sec:gate}.

\section{Device and interactions \label{sec:device}}
To demonstrate the CPHASE gate in experiment, we use two P-mons (P1 and P2) connected via a fixed-frequency resonator as shown in Fig.~\ref{fig:circuit}(c). 
The device is fabricated on a high-Ohmic silicon
substrate with niobium ground planes and $\text{Al/AlO}_x/\text{Al}$
Josephson junctions. To shift the parasitic mode $\mathcal{C}$ to a high frequency, circuit capacitances are designed such that $C_{0i} \ll C_{ij}$ for $i,j \in \{1, 2, 3\}$. This places modes $\mathcal{A}$ and $\mathcal{B}$ at frequencies between $4.5$ and $7.1$~GHz, while mode $\mathcal{C}$ is shifted above $25$~GHz. 
To ensure charge noise protection, the modes $\mathcal{A}$ and $\mathcal{B}$ are operated in the transmon
regime \cite{Koch2007, Gambetta2011, Wills2022, Pfeiffer2026}. We use an on-chip flux line to control the magnetic flux $\Phi_{\text{ext}}$ threading the SQUID loop for frequency tuning of mode $\mathcal{B}$. For XY-control, a charge line is capacitively coupled to island 2 to enable microwave driving of both the qubit and mediator modes.

We measure the P-mons by coupling them to additional readout resonators at frequencies $\omega_{\text{r}1}/2\pi = 6.798$~GHz and $\omega_{\text{r}2}/2\pi = 6.956$~GHz, respectively, using a cross-Kerr readout scheme \cite{Dassonneville2020} which will be described in detail in a forthcoming publication~\cite{Pfeiffer2026}.

Finally, a $\lambda/2$-coplanar waveguide bus resonator with frequency $\omega_r / 2\pi = 8.143~\textrm{GHz}$ mediates the interaction between the two P-mons. By capacitively coupling the bus-resonator to circuit island 3 of each P-mon, we establish the interaction between the mediator modes. 

\begin{figure}[H]
    \centering
    \includegraphics{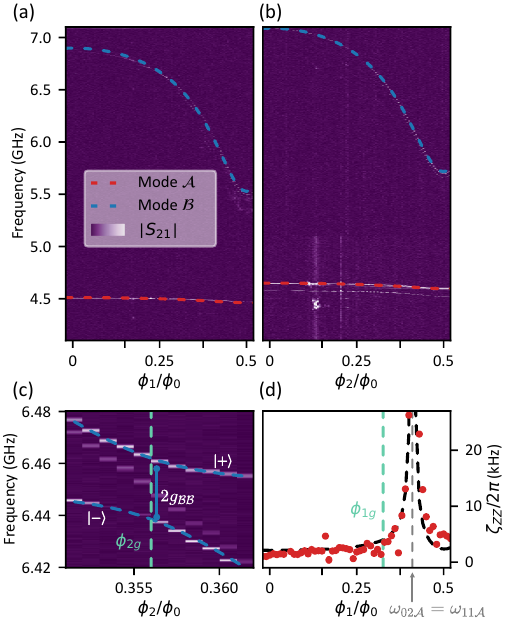}      
 \caption{Spectrum of P-mon P1 (a) and P2 (b) vs. applied magnetic flux $\phi_{1/2}/\phi_0$. 
 (c) Mediator spectrum of P2 vs applied flux $\phi_2/\phi_0$ with P1 parked at the gate operation point $\phi_{1g}$, showing the avoided crossing between the mediator modes of P1 and P2. The blue dashed line indicates a fit with the upper and lower branch labeled by the states $|+\rangle$  and $|-\rangle$, yielding a mediator-mediator coupling $g_{\mathcal{BB}}/2\pi = 11.7(2)$~MHz. The dashed green line indicates the gate operation point $\phi_{2g}$.
 (d) $ZZ$ interaction $\zeta_{\rm ZZ}$ between the qubit modes of P1 and P2 in dependence of $\phi_1/\phi_0$ with P2 parked at the gate operation point $\phi_{2g}$. Error bars are smaller than the data points. The dashed black line is a fit to Eq.~(\ref{eq:zz_qubits}), resulting in a residual qubit-qubit coupling $g_{\mathcal{AA}}/2\pi = 136(1)$~kHz. The dashed grey line indicates the flux point where the states $|1,1\rangle_\mathcal{A}$ and $|0,2\rangle_\mathcal{A}$ become degenerate, while the dashed green line indicates the gate operation point $\phi_{1g}$.}
\label{fig:spectrum}
\end{figure}

To characterize P1 and P2, we measure their transition frequencies as a function of magnetic flux using two-tone spectroscopy. The resulting spectra, shown in Fig.~\ref{fig:spectrum}(a,b), reveal qubit mode transition frequencies $\omega_{\mathcal{A},i}/2\pi$ around $4.5$~GHz and $4.6$~GHz. As intended, these qubit modes [dashed red lines in Fig.~\ref{fig:spectrum}(a,b)] exhibit only a weak flux dependence due to weak hybridization. In contrast, the mediator mode frequencies $\omega_{\mathcal{B},i}/2\pi$ are tunable over ranges of $5.5$–$6.9$~GHz and $5.5$–$7.1$~GHz (dashed blue lines in Fig.~\ref{fig:spectrum}(a,b)), respectively. The measured spectra are in excellent agreement with the circuit model fits. These fits include the measured self-Kerr interactions of the qubit and mediator modes, with further details on the extracted circuit parameters provided in Appendix~\ref{app:circuit_parameters}.

To characterize the bus-mediated interaction, we fix the flux of the first P-mon at $\phi_1/\phi_0 = 0.326$ and sweep the flux $\phi_2/\phi_0$ to bring the two mediator modes into resonance. This resonance occurs at $(\phi_{1} / \phi_0, \phi_{2} / \phi_0) = (0.326, 0.356)$, which we define as the gate operation point $(\phi_{1g}, \phi_{2g})$. The measured spectrum displays a clear avoided crossing [Fig.~\ref{fig:spectrum}(c)], which we fit to the frequencies of two hybridizing modes, $\omega_\pm = \frac{1}{2} \left[ (\omega_1 + \omega_2) \pm \sqrt{(\omega_1 - \omega_2)^2 + 4g_{\mathcal{BB}}^2} \right]$. In this fit, we assume that the second mediator frequency $\omega_2 = \omega_{2,0} + s \cdot \phi_2$ varies linearly with the flux bias over the measured tuning range. From the resulting fit, we extract an effective mediator-mediator coupling strength of $g_{\mathcal{BB}}/2\pi = 11.7(2)$~MHz. Note that the faint spectral line visible between the avoided crossing branches arises from a thermal population of the qubit mode in P1. The resulting cross-Kerr shift moves the P1 mediator transition out of resonance, leaving the uncoupled transition of P2's mediator mode visible at its bare frequency.

Device imperfections, such as capacitance or Josephson junction asymmetries, can lead to unwanted hybridization of modes $\mathcal{A}$ and $\mathcal{B}$. This hybridization induces a residual direct linear qubit-qubit coupling of the form $g_\mathcal{AA} (\hat{a}_{\mathcal{A}, 1}^\dagger \hat{a}_{\mathcal{A}, 2} + \text{h.c.})$ (see Appendix~\ref{app:effective_hamiltonian} for details). To evaluate the impact of this coupling, we characterize the consequent $ZZ$-type crosstalk, $\zeta_{\text{ZZ}}$, by using a Ramsey sequence to measure the frequency shift of qubit $\mathcal{A}_1$ conditioned on state of qubit $\mathcal{A}_2$. By parking P2 at $\phi_{2g}$, and sweeping $\phi_1$, we obtain $\zeta_{\text{ZZ}}$ as a function of flux, as shown in Fig.~\ref{fig:spectrum}(d). From a fit to \cite{DiCarlo2009, Barends2014}
\begin{align}
    \zeta_{ZZ} = 2 g_{\mathcal{AA}}^2 \left(  \frac{\alpha_{\mathcal{A},1} + \alpha_{\mathcal{A},2}}{(\Delta_{12} + \alpha_{\mathcal{A},1})(\Delta_{12} - \alpha_{\mathcal{A},2})}  \right)  \label{eq:zz_qubits},
\end{align}
\noindent where $\Delta_{12} = \omega_{\mathcal{A},1} - \omega_{\mathcal{A},2}$ is the qubit-qubit detuning, we extract a residual direct qubit-qubit interaction of $g_{\mathcal{AA}}/2\pi = 136(1)$~kHz. $\zeta_{\text{ZZ}}$ increases rapidly near the resonance between the states $|0,2\rangle_\mathcal{A}$ and $|1,1\rangle_\mathcal{A}$, resulting in the divergence predicted by Eq.~\eqref{eq:zz_qubits}. However, in regions where these levels are sufficiently detuned, $\zeta_{\text{ZZ}}$ is strongly suppressed. At the gate operation point $\phi_{1g}$, the residual interaction is $\zeta_{\text{ZZ}} = 3.6(5)~\textrm{kHz}$, demonstrating that the computational modes remain isolated, even when the mediators are configured for maximum interaction. This is further confirmed by simultaneous randomized benchmarking \cite{Gambetta2012} of single-qubit gates (see Appendix~\ref{app:single_qubit gates}). For qubit mode $\mathcal{A}_1$ gate fidelities for individual ($\mathcal{F}_{I}$) versus simultaneous ($\mathcal{F}_{S}$) operation are $\mathcal{F}_{I} = 99.971(2)\%$ and $\mathcal{F}_{S}=99.963(1)\%$, while for qubit mode $\mathcal{A}_2$ we measure $\mathcal{F}_{I}=99.950(1)\%$ and $\mathcal{F}_{S}=99.944(1)\%$. These comparable results demonstrate minimal performance degradation during simultaneous addressing.

\section{Mediator-activated CPHASE gate \label{sec:gate}}
To realize a CPHASE gate the flux bias points are set to the gate operation point $(\phi_{1g}, \phi_{2g}) = (0.326, 0.356)$ 
\begin{figure}[H]
    \centering
    \includegraphics{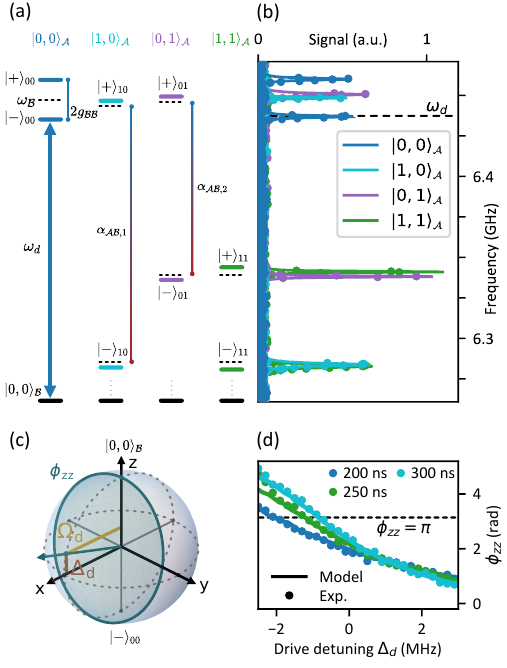}      
 \caption{(a) Energy levels of coupled mediator modes depending on the qubit states $|i,j\rangle_\mathcal{A}$. Dashed horizontal lines indicate bare mediator energy levels in the absence of coupling $g_\mathcal{BB}$ (blue dot-ended arrow); these are shifted by cross-Kerr interactions $\alpha_{\mathcal{AB},1/2}$ (blue/red dot-ended arrows). The blue arrow marks the target transition $|0,0\rangle_\mathcal{B} \leftrightarrow |-\rangle_{00}$ for the gate. (b) Two-tone spectroscopy of the mediator spectrum for different initial qubit states $|i,j\rangle_\mathcal{A}$, where the dashed line indicates the drive frequency $\omega_d$ used for the CPHASE gate. (c) Bloch sphere of the $\{|00\rangle_\mathcal{B}, |-\rangle_{00} \}$ subspace. The green area indicates the cyclic trajectory yielding a geometric phase $\phi_{\rm ZZ}$, determined by the drive detuning $\Delta_d$ and amplitude $\Omega_d$. (d) Entangling phase $\phi_{\rm ZZ}$ vs. drive detuning $\Delta_d$ for various pulse lengths. Solid lines indicate the simulated phase (see Appendix~\ref{app:numerical_simulations}).}
\label{fig:cphase_mechanism}
\end{figure}
\noindent (see Sec.~\ref{sec:device}). At this point, the bare mediator frequencies become resonant, $\omega_{\mathcal{B}, 1} = \omega_{\mathcal{B}, 2} \equiv \omega_{\mathcal{B}}$. 
The cross-Kerr interaction shifts the frequencies to $\tilde{\omega}_{\mathcal{B}, 1, i} = \omega_{\mathcal{B}} + i \alpha_{\mathcal{A}\mathcal{B}, 1}$ and $\tilde{\omega}_{\mathcal{B}, 2, j} = \omega_{\mathcal{B}} + j \alpha_{\mathcal{A}\mathcal{B}, 2}$ depending on the qubit state $|i,j\rangle_\mathcal{A}$.
Through the interaction $g_\mathcal{BB}$, the mediators hybridize with eigenstates $|+\rangle_{ij} = \sin(\theta_{ij}) |1,0\rangle_\mathcal{B} + \cos(\theta_{ij}) |0,1\rangle_\mathcal{B}$ and $|-\rangle_{ij} = \cos(\theta_{ij}) |1,0\rangle_\mathcal{B} - \sin(\theta_{ij}) |0,1\rangle_\mathcal{B}$ and a hybridization angle defined by $\tan(2\theta_{ij}) = 2g_{\mathcal{BB}} / \Delta_{ij}$. This results in the transition frequencies $\omega_{\pm, ij} = (\Sigma_{ij} \pm \sqrt{\Delta_{ij}^2 + 4 g_\mathcal{BB}^2})/2$, where $\Sigma_{ij} = \tilde{\omega}_{\mathcal{B}, 1, i} + \tilde{\omega}_{\mathcal{B}, 2, j}$ and $\Delta_{ij} = \tilde{\omega}_{\mathcal{B}, 1, i} - \tilde{\omega}_{\mathcal{B}, 2, j}$ represent the state-dependent sum and difference frequencies, respectively. This spectrum is sketched in Fig.~\ref{fig:cphase_mechanism}(a), illustrating how the cross-Kerr interactions shift the respective mediator frequency and modify the effective hybridization for each qubit configuration. 

The mediator transition frequencies $\omega_{\pm,ij}$ are measured via two-tone spectroscopy by preparing the system in each of the four computational states $|ij\rangle_\mathcal{A}$. The recorded spectra shown in Fig.~\ref{fig:cphase_mechanism}(b) reveal distinct peaks that align closely with the frequencies predicted by our model using independently measured values for the coupling $g_\mathcal{BB}$ and cross-Kerr strengths $\alpha_{\mathcal{AB}, 1/2}$ (see Tab.~\ref{tab:qubit_params_phi_g}).

To eventually implement a CPHASE gate, we drive the $|0,0\rangle_\mathcal{B} \leftrightarrow |-\rangle_{00}$ transition, as it is maximally detuned from all other transitions. A cyclic $2\pi$-evolution results in a geometric phase $\phi_{00}$ determined by the area enclosed by the trajectory on the Bloch sphere $\{ |0,0\rangle_\mathcal{B}, |-\rangle_{00}\}$, as illustrated in Fig.~\ref{fig:cphase_mechanism}(c). While a resonant drive yields the maximum phase $\phi_{00} = \pi$, a finite drive detuning $\Delta_d = \omega_{-,00} - \omega_d$ serves as a control knob for $\phi_{00}$. Ideally, the drive only affects the target transition, however, off-resonant ``non-target'' transitions must be considered if the pulse is short compared to the drive detuning. The closest non-target transitions are $|0,0\rangle_\mathcal{B} \leftrightarrow |+\rangle_{10}$ and $|0,0\rangle_\mathcal{B} \leftrightarrow |+\rangle_{01}$, with a detuning of approximately $g_\mathcal{BB}$ in the limit $g_\mathcal{BB} \ll \alpha_{\mathcal{AB},i}$.
The detuned drive on these transitions does not degrade the gate performance as long as the state trajectories leave no residual population in the mediator modes after the gate. This condition is satisfied when the non-target transitions undergo adiabatic evolution, simply contributing additional state-dependent phases $\phi_{ij}$ to the total unitary. Note that single-qubit phases can be compensated cost-free via virtual Z-gates \cite{McKay2016}. For each qubit state $|i,j\rangle_\mathcal{A}$, the evolution then satisfies $|ij\rangle_\mathcal{A} \otimes |0,0\rangle_\mathcal{B} \rightarrow e^{i\phi_{ij}} |ij\rangle_\mathcal{A} \otimes |0,0\rangle_\mathcal{B}$, and the resulting entangling phase is given by $\phi_{ZZ} = \phi_{11} + \phi_{00} - \phi_{01} - \phi_{10}$.

Since the requirement for adiabatic evolution of non-target transitions imposes a speed limit on the gate, we employ two strategies to enhance drive selectivity and thereby enable faster operation. First, we simultaneously drive via both drive lines and adjust the relative phase to exploit the state's symmetry \cite{Filipp2011, Fillip2011_2}. With the drives in-phase, the target state $|-\rangle_{00}$ becomes a ``bright state'', enhancing the Rabi rate by a factor of $\sqrt{2}$ relative to the closest non-target states, $|+\rangle_{10}$ and $|+\rangle_{01}$ (see Appendix~\ref{app:mediator_dynamics}).
The non-target symmetric state $|+\rangle_{00}$ then remains a ``dark state''. Second, advanced pulse shaping techniques are used to further suppress spectral weight of the pulse at non-target frequencies \cite{Hyyppa2024, Singh2026} for the gate implementation, as discussed in Sec.~\ref{sec:benchmarking}.

To experimentally quantify the entangling phase, we perform a Ramsey-style sequence \cite{Ganzhorn2020} using simple Gaussian pulses of three different lengths. At each drive detuning $\Delta_d$, we calibrate the drive amplitude to ensure a cyclic $2\pi$ mediator evolution (see Appendix~\ref{app:calibration}). As shown in Fig.~\ref{fig:cphase_mechanism}(d), $\Delta_d$ acts as a precise control knob for $\phi_{ZZ}$. We find that shorter pulse durations lead to different offsets in the phase accumulation due to the broader spectral weight of the drive increasing the contribution from non-target transitions. These results are well-captured by a numerical simulation using QuTiP   \cite{qutip5} (see Appendix~\ref{app:numerical_simulations}). The simulation, which includes no free parameters and accounts for mediator self-Kerr and higher excited states, shows excellent agreement with the experimental data across all pulse durations. 

Both experiment and simulation demonstrate that the drive detuning $\Delta_d$ serves as a calibration parameter for the entangling phase $\phi_{ZZ}$, which can be set to $\phi_{ZZ}=\pi$ [dashed line in Fig.~\ref{fig:cphase_mechanism}(d)] to realize a CZ gate.

\section{CZ-Gate Performance \label{sec:benchmarking}}
To evaluate the performance of the CZ gate, we perform interleaved randomized benchmarking (IRB) \cite{Magesan2012} for varying pulse durations. We compare two pulse shaping strategies: Gaussian envelopes and a Gaussian higher-derivative DRAG (HD-DRAG) scheme \cite{Hyyppa2024} to avoid excitation of non-target states.

\begin{figure}[H]
    \centering
    \includegraphics{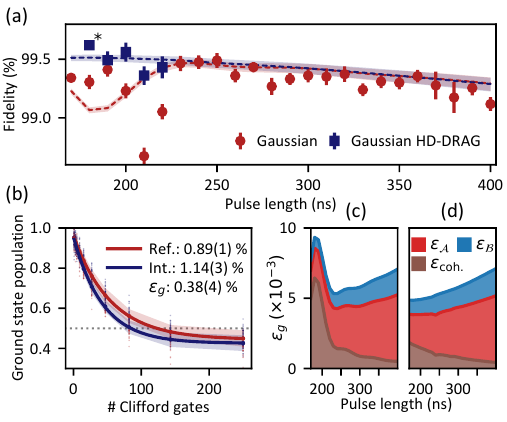}      
 \caption{(a) Gate fidelity vs. pulse length for Gaussian and HD-DRAG pulses. Dashed lines denote mean simulated fidelity, shaded regions represent simulated fidelity bounds derived from the standard deviation of time-averaged coherence times. (b) Interleaved randomized benchmarking corresponding to the data point marked by (*) in (a). The grey dashed line indicates 0.5 ground state population. (c) Simulated error budget for Gaussian pulses vs pulse length. The different colors indicate coherent errors $\epsilon_{\rm coh.}$ (brown), qubit mode decay and dephasing $\varepsilon_\mathcal{A}$ (red) and mediator mode decay $\varepsilon_\mathcal{B}$ (blue). (d) Simulated error budget for Gaussian HD-DRAG pulses.}
\label{fig:cphase_benchmarking}
\end{figure}

We first characterize the gate performance using a Gaussian envelope $g(t) = \exp[-(t-t_p/2)^2/2\sigma^2]$ with a fixed duration-to-width ratio of $t_p = 4\sigma$. We calibrate the gate for durations from $170$ to $400~\text{ns}$ (see Appendix~\ref{app:calibration}) and observe a maximum gate fidelity of $\mathcal{F} = 99.49(6)\%$ at a duration of $250~\text{ns}$. For pulses exceeding this duration, the fidelity decreases as the pulse length increases [see Fig.~\ref{fig:cphase_benchmarking}(a)], indicating a coherence-limited regime. Conversely, reducing the duration below $230~\text{ns}$ leads to a sharp decrease in performance with a prominent dip at $210~\text{ns}$. Interestingly, the fidelity partially recovers for pulses shorter than $200~\text{ns}$, though it remains below the $250~\text{ns}$ peak. This sharp dip and subsequent recovery suggest that the gate is dominated by coherent error mechanisms rather than decoherence.

To obtain a better understanding, we analyze the error budget using numerical simulations in QuTiP  \cite{qutip5} (see Appendix~\ref{app:numerical_simulations}). Our model incorporates coherent errors ($\varepsilon_{\rm coh.}$), qubit decay and dephasing ($\varepsilon_{\mathcal{A}}$), as well as mediator decay ($\varepsilon_{\mathcal{B}}$) with the corresponding coherence times given in Tab.~\ref{tab:coherence_times_gate}. Notably, the simulations align most accurately with the IRB data when mediator dephasing is excluded, suggesting the gate protocol is robust to the slow $1/f$ flux-noise typically present in SQUID-tunable circuits \cite{Yoshihara2006, Kauyanagi2007, Bylander2011, Hutchings2017, Pfeiffer2024}. The resulting simulated fidelities, shown in Fig.~\ref{fig:cphase_benchmarking}(c) are compared to the measurements in Fig.~\ref{fig:cphase_benchmarking}(a). The simulation shows that the incoherent errors $\varepsilon_{\mathcal{A}}$ and $\varepsilon_{\mathcal{B}}$ increase with gate time, accurately capturing the fidelity trend for long pulse durations. At shorter durations the model reveals a sharp increase in the coherent error $\varepsilon_{\rm coh.}$ caused by residual mediator population after the gate (see Appendix~\ref{app:numerical_simulations}). This leakage stems from the increased spectral width of the pulses, which leads to off-resonant driving of non-target transitions, with the largest contributions from the transitions $|0,0\rangle_\mathcal{B} \leftrightarrow |+\rangle_{10}$ and $|0,0\rangle_\mathcal{B} \leftrightarrow |+\rangle_{01}$. Consistent with measurements, the simulation shows a fidelity recovery at the shortest gate durations. This behavior stems from the frequency spectrum of the truncated Gaussian pulses: at specific lengths, spectral gaps align with off-resonant transitions, suppressing unwanted excitations despite the broader bandwidth. The fact that both the measured dip and recovery occur at slightly longer pulse durations than simulated is likely due to experimental pulse distortions in the control lines.

To systematically mitigate the spectral overlap with non-target transitions, we employ a HD-DRAG pulse \cite{Hyyppa2024}. This pulse utilizes higher-order derivatives of the Gaussian envelope to engineer the pulse spectrum, creating spectral gaps at the frequencies of the nearest non-target transitions (see Appendix~\ref{app:calibration}).

We characterize the HD-DRAG pulses for durations between $180$ and $220~\text{ns}$, focusing on the region where the standard Gaussian pulses exhibited the most significant coherent errors. The lower bound of this range is determined by experimentally accessible drive power. As shown in Fig.~\ref{fig:cphase_benchmarking}(a), the HD-DRAG pulses maintain high fidelity throughout this window, successfully bypassing the fidelity dip observed with standard Gaussian pulses. Our highest measured performance is achieved with a $180~\text{ns}$ HD-DRAG pulse, yielding a fidelity of $\mathcal{F} = 99.62(4)\%$. The corresponding IRB curve is shown in Fig.~\ref{fig:cphase_benchmarking}(b). For these measurements, we extract the gate fidelity by monitoring only the population of mode $\mathcal{A}_1$, which is sufficient for randomized benchmarking of a two-qubit gate \cite{McKay2019, Ganzhorn2020} and avoids complications from flux-crosstalk between flux pulses used during the readout of both qubits. While the ground-state population is expected to converge to 0.5 on average for long Clifford sequences, the visible deviation indicates that even for HD-DRAG pulses, coherent errors and residual leakage into mediator states become an issue at shorter pulse durations.

We further analyze the HD-DRAG performance using the same numerical model and error sources as used in 
the Gaussian case. The simulated error budget [Fig.~\ref{fig:cphase_benchmarking}(d)] confirms that HD-DRAG pulses significantly reduce the coherent error contribution $\varepsilon_{\rm coh.}$ compared to the Gaussian pulses. At these short gate times, $\varepsilon_{\rm coh.}$ is suppressed to a level comparable with the incoherent errors $\varepsilon_{\mathcal{A}}$ and $\varepsilon_{\mathcal{B}}$, yielding the overall improved gate fidelity.

\section{Conclusion}
We have extended the P-mon's protection mechanism against coupling induced decoherence channels to two-qubit interactions, which was originally designed to mitigate decoherence channels induced by the readout resonator.
By implementing a mode-selective coupling between two P-mons, we show that unwanted $ZZ$ crosstalk is suppressed while operating a CPHASE gate based on qubit-state-dependent mediator dynamics. 
Using HD-DRAG pulses, with spectral gaps at non-target transitions, we realize a CZ gate with a fidelity of up to $\mathcal{F}=99.62(4)\%$, as benchmarked via interleaved randomized benchmarking.

To further inhibit residual $ZZ$ crosstalk in future devices, qubit-mediator hybridization can be reduced through advanced Josephson junction parameter targeting \cite{Osman2021, Hertzberg2021} which minimizes circuit asymmetries. Furthermore, qubit frequency targeting can be improved to avoid collisions between the $|11\rangle_\mathcal{A}$ state and the $|02\rangle_\mathcal{A}$ or $|20\rangle_\mathcal{A}$ manifolds, thereby suppressing parasitic $ZZ$ interactions as described in Eq.~(\ref{eq:zz_qubits}). Because the gate scheme is fundamentally independent of specific qubit frequencies, such targeting is feasible. Notably, this flexibility also allows to circumvent frequency crowding and the associated microwave crosstalk. 

To further enhance the gate fidelities, one can increase the mediator-mediator coupling $g_\mathcal{BB}$, which creates larger detunings from non-target transitions and enables thus faster gate operations. Additionally, errors due to mediator decay can be reduced by operating at a frequency further detuned from the readout resonators to minimize Purcell loss \cite{Houck2008}. While the dependence of gate error on mediator dephasing requires further investigation, dephasing times can be optimized by operating at the first-order flux-insensitive point ($\phi/\phi_0 = 0$), which requires more precise frequency targeting.

Looking ahead, multiple P-mons could be coupled via a shared bus resonator, enabling pairwise gates between any coupled qubits by tuning the relevant mediators into resonance. This approach offers the potential for higher connectivity \cite{Vigneau2025} than a square lattice and could be extended to multi-qubit operations by simultaneously tuning three or more mediator modes into resonance.

In summary, by combining the previously demonstrated protection against readout-induced decoherence \cite{Pfeiffer2024} with the suppression of inter-qubit crosstalk presented here, the P-mon architecture is a compelling candidate for scalable quantum processors. Investigating high-connectivity topologies and multi-qubit gates remains a promising direction for future research.

\section{Acknowledgments}
This research was funded by the BMW Group.
We acknowledge financial support from the European Union’s Horizon 2020 research and innovation program ‘OpenSuperQPlus100' (Grant No.
955479), the BMBF programs ‘German Quantum Computer based on Superconducting Qubits’ (GeQCoS; Nr. 13N15680) and MUNIQCSC (Nr. 13N16188),
the German Research Foundation project ‘Multi-qubit gates for the efficient exploration of Hilbert space with superconducting qubit systems’
(Nr. 445948657) and the excellence initiative ‘Munich
Center for Quantum Science and Technology’ (MCQST;
Nr. 390814868) as well as the Munich Quantum Valley,
which is supported by the Bavarian state government
with funds from the Hightech Agenda Bayern Plus.

\appendix

\section{Effective Hamiltonian \label{app:effective_hamiltonian}}
The circuit Hamiltonian of a single P-mon is given by
\begin{align}
    \hat{H}_{\rm c} =& \frac{1}{2} \sum_{i,j=1}^3 (\boldsymbol{C}^{-1})_{ij} \hat{q}_i \hat{q}_j - \sum_i E_{Ji} \cos\hat{\phi}_i \label{eq:pmon_circuit_hamiltonian}, 
\end{align}
where $i, j \in \{1, 2, 3\}$ index the circuit islands. The operators $\hat{\phi}_i$ and $\hat{q}_i$ represent the phase and charge on island $i$ relative to the reference node (island 0). The parameters include the capacitance matrix $\boldsymbol{C}$, individual Josephson energies $E_{J1}, E_{J2}$, and the effective SQUID energy $E_{J3}(\phi_{\textrm{ext}})$.
Here, $C_{ij}$ are the effective capacitances, obtained by accounting for ground capacitances $C_{gi}$ and eliminating the static mode.

Coupling the P-mon to a resonator of frequency $\omega_r$ is described by the total Hamiltonian
\begin{align}
\hat{H}_{\rm Pr} = \hat{H}_{\rm c} + \omega_r \hat{a}_r^\dagger \hat{a}_r + \sum_{k=1}^3 (\boldsymbol{C}^{-1})_{kr} \hat{q}_k \hat{q}_r,\label{eq:pmon_resonator_circuit_hamiltonian}
\end{align}
where $\hat{a}_r$ is the resonator annihilation operator,  $\hat{q}_r = i \sqrt{\hbar / 2Z_r}(\hat{a}_r^\dagger - \hat{a}_r)$ the resonator charge operator, and $Z_r$ the resonator impedance. Following Ref.~\cite{Pfeiffer2024}, we transform the Hamiltonian into the normal mode basis of the P-mon and expand the potential to fourth order to obtain the effective Hamiltonian 
\begin{align}
\hat{H}_{\rm Pr} &= \sum_m \omega_m \hat{a}_m^\dagger \hat{a}_m + \sum_m
\dfrac{\alpha_m}{2} \hat{a}_m^\dagger \hat{a}_m^\dagger \hat{a}_m \hat{a}_m \nonumber \\
&+ \sum_{m\neq n} g_{mn} (\hat{a}_m^\dagger \hat{a}_n + \hat{a}_m \hat{a}_n^\dagger) + \alpha_{mn} \hat{a}_m^\dagger \hat{a}_m \hat{a}_n^\dagger \hat{a}_n \nonumber \\
&+  \omega_r \hat{a}_r^\dagger \hat{a}_r + \sum_m g_{mr} (\hat{a}^\dagger_r \hat{a}_m + \hat{a}_r \hat{a}_m^\dagger),
\label{eq:pmon_effective_hamiltonian}
\end{align}
with mode frequencies $\omega_m$, self-Kerr interactions $\alpha_m$, cross-Kerr interactions $\alpha_{mn}$, mode-mode couplings $g_{mn}$, and resonator-mode couplings $g_{mr}$ for the modes $m, n \in \{\mathcal{A}, \mathcal{B}, \mathcal{C}\}$. Since mode $\mathcal{C}$ is designed to have a high frequency ($\omega_\mathcal{C} > 25$~GHz) such that $\omega_\mathcal{C} \gg \omega_\mathcal{A}, \omega_\mathcal{B}$, we neglect it in the following analysis. The mode-mode coupling $g_{\mathcal{AB}}$ arises exclusively from circuit asymmetries; specifically, $g_{\mathcal{AB}} \propto E_{J1} - E_{J2}$ and $g_{\mathcal{AB}} \propto C_{13} - C_{23}$. 
To minimize this coupling, we design the circuit with symmetric capacitances ($C_{13} = C_{23}$) and Josephson energies ($E_{J1} = E_{J2}$). Furthermore, by introducing a large coupling capacitance $C_{3r}$ between the P-mon and the resonator while keeping $C_{1r} = C_{2r}$, the coupling $g_{\mathcal{B}r}$ is established while the qubit mode $\mathcal{A}$ remains decoupled ($g_{\mathcal{A}r} = 0$) \cite{Pfeiffer2024}. Assuming perfect symmetry ($g_{\mathcal{AB}}=0$), we arrive at Eq.~(\ref{eq:pmon_effective_hamiltonian_no_asymmetry}) for a single P-mon and Eq.~(\ref{eq:pmons_bus_hamiltonian}) for two P-mons coupled via a bus resonator.  

We now consider the impact of a finite fabrication asymmetry ($g_{\mathcal{AB}}\neq 0$). 
In this case, modes $\mathcal{A}$ and $\mathcal{B}$ hybridize into the dressed modes $\mathcal{A'}$ and $\mathcal{B'}$ via the transformation
\begin{align}
    \hat{a}_\mathcal{A'} &= \cos\theta_\mathcal{AB} \hat{a}_\mathcal{A} + \sin\theta_\mathcal{AB} \hat{a}_\mathcal{B}, \nonumber \\
    \hat{a}_\mathcal{B'} &= -\sin\theta_\mathcal{AB} \hat{a}_\mathcal{A} + \cos\theta_\mathcal{AB} \hat{a}_\mathcal{B},
\end{align}
where the hybridization angle is given by $\tan(2\theta_\mathcal{AB}) = 2g_\mathcal{AB}/\Delta_\mathcal{AB}$  with detuning $\Delta_\mathcal{AB} = \omega_\mathcal{B} - \omega_\mathcal{A}$.
Consequently, the interaction Hamiltonian $\hat{H}_{\textrm{int}} = g_{\mathcal{B}r} (\hat{a}_\mathcal{B}^\dagger \hat{a}_r + \textrm{h.c.})$ transforms to
\begin{align}
    \hat{H}_{\textrm{int}} = g_{\mathcal{B}r} \left(\cos\theta_\mathcal{AB} \hat{a}_\mathcal{B'}^\dagger \hat{a}_r + \sin\theta_\mathcal{AB} \hat{a}_\mathcal{A'}^\dagger \hat{a}_r + \textrm{h.c.}\right),
\end{align}
which reveals that the dressed mode $\mathcal{A'}$ now also couples to the resonator with an effective strength $g_{\mathcal{A'}r} = \sin(\theta_\mathcal{AB}) g_{\mathcal{B}r}$. 

For two P-mons (indexed $\mu \in \{1,2\}$) featuring identical hybridization angles $\theta_\mathcal{AB}$ and equal mode-resonator couplings $g_{\mathcal{B}r}$, this induces an unwanted bus-mediated interaction between the two qubit modes:
\begin{align}
    g_\mathcal{A'A'} = \sin^2\theta_\mathcal{AB} \, g_{\mathcal{B}r}^2 \left(\frac{1}{\Delta_{\mathcal{A'}r,1}} + \frac{1}{\Delta_{\mathcal{A'}r,2}}\right),
\end{align}
where $\Delta_{\mathcal{A'}r,\mu} = \omega_{\mathcal{A'}\mu} - \omega_r$ is the detuning of the $\mu$-th qubit mode $\mathcal{A'}$ from the bus resonator. In the limit of small asymmetries ($\theta_\mathcal{AB} \ll 1$), the hybridization angle scales as $\theta_\mathcal{AB} \approx g_\mathcal{AB} / \Delta_\mathcal{AB}$, yielding an effective qubit-qubit coupling of $g_\mathcal{A'A'} \propto (g_\mathcal{AB} / \Delta_\mathcal{AB})^2$.

This highlights that the residual interaction $g_{\mathcal{A'A'}}$ can be systematically suppressed in two ways. First, one can reduce the hybridization of modes $\mathcal{A}$ and $\mathcal{B}$ by either minimizing fabrication asymmetries to lower $g_{\mathcal{AB}}$ or by increasing their bare mode detuning $\Delta_{\mathcal{AB}}$. Second, by operating at a larger detuning $\Delta_{\mathcal{A'}r,\mu}$ between the qubit modes and the bus resonator to weaken the effective bus-mediated coupling.

\begin{figure}[H]
    \centering
        \includegraphics{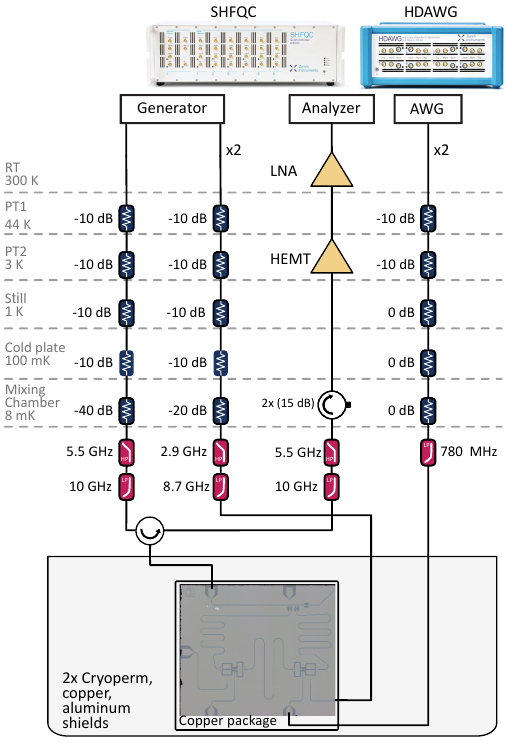}
 \caption{Wiring diagram of the experimental setup.}
 \label{fig:exp_setup}
\end{figure}
\section{Experimental setup \label{app:setup}}
The wiring diagram of the experimental setup is shown in Fig.~\ref{fig:exp_setup}. Measurements are performed in a BlueFors XLD1000 dilution refrigerator with a base temperature of 8 mK. A common input microwave line is used for readout of both P-mons. Individual microwave lines are used for qubit control. Microwave pulses are generated using a Zurich Instruments SHFQC qubit controller. The on-chip flux line is controlled via a Zurich Instruments HDAWG arbitrary waveform generator. Microwave attenuators and filters provide isolation from thermal noise and noise from the room temperature signal sources. The readout signal is amplified using a LNF-LNC4\_8F high electron mobility transistor (HEMT) and BZ-04000800-081045-152020 low noise amplifier (LNA). Single shot readout is performed by the Zurich Instruments SHFQC waveform generator and analyzer. 

\section{Circuit parameters and coherence times \label{app:circuit_parameters}}
We obtain the circuit parameters (Tab.~\ref{tab:circuit_parameters}) by fitting both the measured spectrum (Fig.~\ref{fig:spectrum}) and the self-Kerr interactions of the qubit and mediator modes to the circuit Hamiltonian defined in Eq.~(\ref{eq:pmon_circuit_hamiltonian}) using numerical diagonalization in the charge basis and a Nelder-Mead optimization \cite{SciPy2020}. To reduce the parameter space, we enforce two levels of constraints. First, we assume identical capacitance matrices for both P-mon 1 and P-mon 2, which is a reasonable approximation given that they are equal in the design and the spread in fabricated capacitances is negligible. Second, we impose the designed symmetry constraints ($C_{13}=C_{23}$, $C_{g1} = C_{g2}$, and $C_{01}=C_{02}$), which further reduces the number of independent variables.
\begin{table}[H]
\caption{\label{tab:circuit_parameters}
Fitted circuit parameters for P-mon 1 (P1) and P-mon 2 (P2): capacitances $C_{ij}$ between the circuit islands $i=0, 1, 2, 3$ and the ground (g), Josephson energies $E_{Ji}$ and SQUID asymmetry $d_{\rm SQUID}.$}
\begin{ruledtabular}
\begin{tabular}{cccccc}
$C_{ij}$ (fF) & g & 0 & 1 & 2 & 3\\
\colrule
g & - & 8.6 & 60.0 & 60.0 & 121.7 \\
0 & 8.6 & - & 2.3 & 2.3 & 2.8 \\
1 & 60.0 & 2.3 & - & 17.5 & 2.0 \\
2 & 60.0 & 2.3 & 17.5 & - & 2.0 \\
3 & 121.7 & 2.8 & 2.0 & 2.0 & -  
\end{tabular}
\begin{tabular}{ccccc}
 & $E_{J1}/2\pi$ (GHz) & $E_{J2}/2\pi$ (GHz) & $E_{J3}/2\pi$ (GHz) &  $d_{\rm SQUID}$ \\
 P1 & 15.3 & 15.3 & 117.3 & 0.26  \\
 P2 & 15.0 & 17.9 & 109.8 & 0.29  
\end{tabular}

\end{ruledtabular}
\end{table}
\begin{table}[H]
\caption{\label{tab:qubit_params_phi_g}%
Parameters for modes $m \in \{\mathcal{A}, \mathcal{B}\}$ of the individual P-mons, measured at flux points $(\phi_1/\phi_0, \phi_2/\phi_0) = (0.326, 0)$ for P-mon 1 and $(0, 0.356)$ for P-mon 2. Mode frequencies $\omega_m$ are measured via Ramsey sequences, whereas self-Kerr terms $\alpha_m$ and cross-Kerr couplings are determined via spectroscopy.}
\begin{ruledtabular}
\begin{tabular}{ccccc}
& $\mathcal{A}_1$ & $\mathcal{A}_2$  & $\mathcal{B}_1$ & $\mathcal{B}_2$\\
\colrule
$\omega_m/2\pi$ (GHz) & 4.499 & 4.633 & 6.448 & 6.449\\
$\alpha_m/2\pi$ (MHz)  & -126 & -143 & -106 & -91\\
$\alpha_\mathcal{AB}/2\pi$ (MHz) & -164 & -108 &  & \\
\end{tabular}
\end{ruledtabular}
\end{table}
At the gate operation point (see Sec.~\ref{sec:device}), the measured mode frequencies $\omega_m$, self-Kerr $\alpha_m$ ($m\in \{\mathcal{A},\mathcal{B\}}$) and cross-Kerr $\alpha_\mathcal{AB}$ interactions of the two P-mons are listed in Tab.~\ref{tab:qubit_params_phi_g}.
These model parameters are used for the simulation of the entangling phase [Fig.~\ref{fig:cphase_mechanism}(d)] and the gate fidelity [Fig.\ref{fig:cphase_benchmarking}(a)]. For the gate fidelity we further use the time-averaged decay and coherences times at the gate operation point, listed in Tab.~\ref{tab:coherence_times_gate}.
\begin{table}[H]
\caption{\label{tab:coherence_times_gate}
Time averaged decay times $T_1$ and coherence times $T_2$ of the qubit modes $\mathcal{A}_1$, $\mathcal{A}_2$ and the hybridized mediator mode $\mathcal{B}_{-}$ (corresponding to the $|0,0\rangle_\mathcal{B} \leftrightarrow |-\rangle_{00}$ transition) at the gate operation point.}
\begin{ruledtabular}
\begin{tabular}{cccccc}
  & $T_1$ ($\mu$s) & $T_2$ ($\mu$s) \\
\colrule
$\mathcal{A}_1$ & 139(18) & 102(11)  \\
$\mathcal{A}_2$ & 141(13) & 78(4)  \\
$\mathcal{B}_{-}$ & 12(1) & 5.2(6)  \\
\end{tabular}
\end{ruledtabular}
\end{table}

\section{Driven Mediator mode dynamics \label{app:mediator_dynamics}}
To understand the response of the coupled mediator modes to a drive, we project the Hamiltonian in Eq.~(\ref{eq:H_AB_TLS}) onto the subspace of a fixed qubit state $|i, j\rangle_\mathcal{A}$.
Including a drive on each mediator mode with phase $\phi_{d, \mu}$ and amplitude $\epsilon_{d, \mu}$, the driven Hamiltonian is then given by
\begin{align}
    \hat{H}_{\mathcal{B}} &= \sum_\mu \tilde{\omega}_{\mathcal{B} \mu, k_\mu} \hat{\sigma}_{\mathcal{B} \mu}^\dagger \hat{\sigma}_{\mathcal{B} \mu} + g_{\mathcal{BB}} ( \hat{\sigma}_{\mathcal{B} 1}^\dagger \hat{\sigma}_{\mathcal{B} 2} + \hat{\sigma}_{\mathcal{B} 2}^\dagger \hat{\sigma}_{\mathcal{B} 1}) \nonumber \\
    &+ \sum_\mu \epsilon_{d, \mu} \left(\hat{\sigma}_{\mathcal{B} \mu}^\dagger e^{-i(\omega_d t + \phi_{d,\mu})} + \hat{\sigma}_{\mathcal{B} \mu} e^{+i(\omega_d t + \phi_{d,\mu})}\right)
    \label{eq:H_mediator_drive_fixed_a},
\end{align}
with the qubit state-dependent frequencies $\tilde{\omega}_{\mathcal{B}1, i} = \omega_{\mathcal{B}1} + i \alpha_{\mathcal{A}\mathcal{B}1} $ and $\tilde{\omega}_{\mathcal{B}2, j} = \omega_{\mathcal{B}2} + j \alpha_{\mathcal{A}\mathcal{B}2}$. In the following, we consider the case of equal drive strengths $\epsilon_{d,1} = \epsilon_{d, 2} \equiv \epsilon_d$ and frequencies $\omega_{d, 1} = \omega_{d, 2} \equiv \omega_d$. Without loss of generality, we set $\phi_{d, 1} = 0$, such that the phase difference between the two drives is given by $\phi_{d, 2} \equiv \phi_d$. In the absence of a drive ($\epsilon_{d} = 0$), the mediator Hamiltonian is transformed to its eigenbasis by
\begin{align}
    \hat{\sigma}_{\theta_{ij}, +} &= \sin(\theta_{ij}) \hat{\sigma}_{\mathcal{B}1} + \cos(\theta_{ij}) \hat{\sigma}_{\mathcal{B}2} \nonumber \\
    \hat{\sigma}_{\theta_{ij}, -} &= \cos(\theta_{ij}) \hat{\sigma}_{\mathcal{B}1} - \sin(\theta_{ij}) \hat{\sigma}_{\mathcal{B}2} \label{eq:H_m_rot_frame},
\end{align}
with the hybridization angle $\theta_{ij}$ given by $\tan(2\theta_{ij}) = 2g_{\mathcal{BB}} / (\tilde{\omega}_{\mathcal{B}1, i} - \tilde{\omega}_{\mathcal{B}2, j})$. Transforming to a frame rotating at the drive frequency $\omega_d$ via the unitary $\hat{U} = \exp(i \omega_d \sum_\mu \hat{\sigma}_{\mathcal{B}\mu}^\dagger \hat{\sigma}_{\mathcal{B}\mu})$ yields the stationary Hamiltonian
\begin{align}
    \hat{H} &=  \sum_{\alpha=\pm} \Delta_{\theta_{ij}, \alpha} \hat{\sigma}_{\theta_{ij}, \alpha}^\dagger \hat{\sigma}_{\theta_{ij}, \alpha} + \epsilon_{d, \theta_{ij} , \alpha} \hat{\sigma}_{\theta_{ij}, \alpha}^\dagger + \epsilon_{d, \theta_{ij} , \alpha}^* \hat{\sigma}_{\theta_{ij}, \alpha}\label{eq:H_hybridized_basis}
\end{align}
Here, the drive detunings $\Delta_{\theta_{ij}, \pm} = \omega_{\theta_{ij}, \pm} - \omega_d$ of the hybridized states and the corresponding drive amplitudes are given by
\begin{align}
    \omega_{\theta_{ij}, \pm} &= \frac{1}{2} \left(\tilde{\omega}_{\mathcal{B}1, i} + \tilde{\omega}_{\mathcal{B}2, j} \pm \sqrt{(\tilde{\omega}_{\mathcal{B}1, i} - \tilde{\omega}_{\mathcal{B}2, j})^2 + 4g_{\mathcal{BB}}^2} \right), \nonumber \\
    \epsilon_{d, \theta_{ij} , +} &= \epsilon_d (\sin(\theta_{ij}) + \cos(\theta_{ij})e^{-i\phi_d}), \nonumber \\
    \epsilon_{d, \theta_{ij} , -} &= \epsilon_d (\cos(\theta_{ij}) - \sin(\theta_{ij})e^{-i\phi_d}). \label{eq:hybridized_formulas} 
\end{align}
This form highlights the possible drive-induced transitions and their corresponding driving strength. In general, the two coupled mediator modes form a four-level system with four possible transitions: (i, ii) $|0,0\rangle_\mathcal{B} \leftrightarrow |\pm\rangle_{ij}$, (iii, iv) $|\pm\rangle_{ij}\leftrightarrow |1,1\rangle_\mathcal{B}$, with $|+\rangle_{ij} = \sin(\theta_{ij}) |1,0\rangle_\mathcal{B} + \cos(\theta_{ij}) |0,1\rangle_\mathcal{B}$ and $|-\rangle_{ij} = \cos(\theta_{ij}) |1,0\rangle_\mathcal{B} - \sin(\theta_{ij}) |0,1\rangle_\mathcal{B}$. The effective drive strength $\epsilon_{d, \theta_{ij} , \pm}$ of each transition depends on the hybridization angle $\theta_{ij}$ and the relative phase of the drive $\phi_d$. 
Note that the indices $\pm$ denote the ordering of the eigenenergies ($\omega_{\theta_{ij}, +} > \omega_{\theta_{ij}, -}$), and do not represent the symmetry of the hybridized states.

Given our bus-mediated coupling $g_{\mathcal{BB}} < 0$, for the qubit state $|0,0\rangle_\mathcal{A}$ the hybridization angle is $\theta_{00} = -\pi/4$ and the state $|-\rangle_{00}$ corresponds to the symmetric superposition, while $|+\rangle_{00}$ is the antisymmetric one. For the CPHASE gate described in Sec.~\ref{sec:gate} and Sec.~\ref{sec:benchmarking}, we thus use an in-phase drive ($\phi_d = 0$), resulting in $\epsilon_{d, \theta_{00} , -} = \sqrt{2} \epsilon_d$ and $\epsilon_{d, \theta_{00}, +} = 0$. Consequently, the state $|-\rangle_{00}$ becomes a ``bright state'' and the $|+\rangle_{00}$ state a ``dark state'' \cite{Filipp2011, Fillip2011_2}, which is not affected by the drive.  If the qubits are in one of the single-excitation states $|0, 1\rangle_\mathcal{A}$ or $|1, 0\rangle_\mathcal{A}$, since $\alpha_{\mathcal{AB}, 1/2}\gg g_\mathcal{BB}$, we have $\theta_{01/10} \approx 0$ and $\epsilon_{d, \theta_{01/10}, +} \approx \epsilon_d$. Therefore, the drive strength on the target transition $|0,0\rangle_\mathcal{B} \leftrightarrow |-\rangle_{00}$ is enhanced by a factor of $\sqrt{2}$ compared to the closest non-target transitions $|0,0\rangle_\mathcal{B} \leftrightarrow |+\rangle_{10}$ and $|0,0\rangle_\mathcal{B} \leftrightarrow |+\rangle_{01}$.

\section{Numerical simulations \label{app:numerical_simulations}}
To numerically simulate the time evolution including decoherence, we extend the Hamiltonian in Eq.~(\ref{eq:H_AB_TLS}) to include higher energy levels of the mediator modes and a drive on each mediator mode
\begin{align}
    \hat{H} &= \sum_{\mu} (\omega_{\mathcal{A}\mu} \hat{\sigma}_{\mathcal{A}\mu}^\dagger \hat{\sigma}_{\mathcal{A}\mu} + [\omega_{\mathcal{B}\mu}  + \alpha_{\mathcal{AB}\mu}\hat{\sigma}_{\mathcal{A}\mu}^\dagger \hat{\sigma}_{\mathcal{A}\mu}]
    \hat{a}_{\mathcal{B}\mu}^\dagger \hat{a}_{\mathcal{B}\mu} \nonumber \\ 
    &+  \alpha_{\mathcal{B}\mu} \hat{a}_{\mathcal{B}\mu}^\dagger \hat{a}_{\mathcal{B}\mu}^\dagger \hat{a}_{\mathcal{B}\mu} \hat{a}_{\mathcal{B}\mu})
    + g_\mathcal{BB} (\hat{a}_{\mathcal{B}1}^\dagger \hat{a}_{\mathcal{B}2} + \hat{a}_{\mathcal{B}1} \hat{a}_{\mathcal{B}2}^\dagger) \nonumber \\ 
    &+ \sum_{\mu} \Omega_{d,\mu}(t)( \hat{a}_{\mathcal{B}\mu}^\dagger +  \hat{a}_{\mathcal{B}\mu}),
\end{align}
where the drive signal is given by $\Omega_{d, \mu}(t) = \Omega_I (t) \cos(\omega t +\phi_{d,\mu}) + \Omega_Q (t) \sin(\omega t + \phi_{d, \mu})$. 
We transform the Hamiltonian into a rotating frame matching the qubit mode frequencies and the mediator drive frequency using the unitary operator $\hat{U} = \exp(i \sum_\mu (\omega_{\mathcal{A}\mu} \hat{\sigma}_{\mathcal{A}\mu}^\dagger \hat{\sigma}_{\mathcal{A}\mu} + \omega_d \hat{a}_{\mathcal{B}\mu}^\dagger \hat{a}_{\mathcal{B}\mu}))$. Under the rotating wave approximation (RWA), this yields the time-independent Hamiltonian:
\begin{align}
    \hat{H} &= \sum_{\mu} ([\Delta_{\mathcal{B}\mu}  + \alpha_{\mathcal{AB}\mu}\hat{\sigma}_{\mathcal{A}\mu}^\dagger \hat{\sigma}_{\mathcal{A}\mu}]
    \hat{a}_{\mathcal{B}\mu}^\dagger \hat{a}_{\mathcal{B}\mu} \\ 
    &+  \alpha_{\mathcal{B}\mu} \hat{a}_{\mathcal{B}\mu}^\dagger \hat{a}_{\mathcal{B}\mu}^\dagger \hat{a}_{\mathcal{B}\mu} \hat{a}_{\mathcal{B}\mu})
    + g_\mathcal{BB} (\hat{a}_{\mathcal{B}1}^\dagger \hat{a}_{\mathcal{B}2} + \hat{a}_{\mathcal{B}1} \hat{a}_{\mathcal{B}2}^\dagger) \nonumber \\ 
    &+ \sum_{\mu} \epsilon_{I,\mu}(t)( \hat{a}_{\mathcal{B}\mu}^\dagger +  \hat{a}_{\mathcal{B}\mu}) + \epsilon_{Q,\mu}(t)( \hat{a}_{\mathcal{B}\mu}^\dagger -  \hat{a}_{\mathcal{B}\mu}), \nonumber \label{eq:H_AB_numerical_sim},
\end{align}
with the drive detuning $\Delta_{\mathcal{B}\mu} = \omega_{\mathcal{B}\mu} - \omega_d$ and the drive envelopes:
\begin{align}
    \epsilon_{I, \mu}(t) &= \frac{1}{2} [\Omega_I (t) \cos(\phi_{d, \mu})  + \Omega_Q (t) \sin(\phi_{d, \mu})]\nonumber \\
    \epsilon_{Q, \mu}(t) &= \frac{1}{2} [-\Omega_I (t) \sin(\phi_{d, \mu})  + \Omega_Q (t) \cos(\phi_{d, \mu})].
\end{align}
We fix the absolute phase reference by setting $\phi_{d,1} = 0$, such that the relative phase is given by $\phi_{d, 2} \equiv \phi_d$.

To capture different error channels, we incorporate the decay and dephasing of the qubit modes alongside the decay of the mediator modes. The open-system time evolution is governed by the Lindblad master equation:
\begin{align}
    \frac{\partial \hat{\rho}}{\partial t} = -\frac{i}{\hbar} [\hat{H}, \hat{\rho}] + \sum_{\nu=1, \phi}\sum_\mu \mathcal{D}[\hat{L}_{\nu, \mathcal{A}_\mu}] + \mathcal{D}[\hat{L}_{1,\mathcal{B}_-}], \label{eq:master_eq}
\end{align}
with Lindblad dissipator $\mathcal{D}[\hat{L}] = \hat{L} \hat{\rho} \hat{L}^\dagger - \frac{1}{2}\{\hat{\rho}, \hat{L}^\dagger \hat{L} \}$.
The collapse operators for qubit decay and dephasing are defined as $\hat{L}_{1, \mathcal{A}1/2} = \sqrt{\Gamma_{1,\mu}} \hat{\sigma}_{\mathcal{A}1/2}$ and $\hat{L}_{\phi, \mathcal{A}1/2} = \sqrt{2\Gamma_{\phi, 1/2}} \hat{\sigma}_{\mathcal{A}1/2}^\dagger \hat{\sigma}_{\mathcal{A}1/2}$.
For the decay of the target symmetric coupled mediator state $|-\rangle_{00}$, we define $\hat{L}_{1, \mathcal{B}_-} = \sqrt{\Gamma_{1,\mathcal{B}_-}} (\hat{a}_{\mathcal{B}1} + \hat{a}_{\mathcal{B}2})/\sqrt{2}$. The respective rates are derived from experimental coherence times listed in Tab.~\ref{tab:coherence_times_gate}.

We evaluate Eq.~(\ref{eq:master_eq}) numerically using QuTiP  \cite{qutip5}. Truncating each mediator mode to three energy levels was found to be sufficient for numerical convergence across all studied parameter regimes.
We trace out the mediator modes from the full density matrix $\hat{\rho} (t)$ to obtain the reduced density matrix $\hat{\rho}_\mathcal{A}(t) = \textrm{tr}_\mathcal{B} \hat{\rho}(t)$.
Since the drive induces no population transfer in the  qubit subspace, each matrix element $|i, j\rangle_\mathcal{A} \langle k, l|_\mathcal{A}$ acquires a complex phase $e^{i\beta_{ij, kl}}$. For an ideal target unitary $\hat{U} = \textrm{diag}(e^{i\phi_{00}}, e^{i\phi_{10}}, e^{i\phi_{01}}, e^{i\phi_{11}})$, this phase maps directly to $\beta_{ij,kl} = \phi_{ij} - \phi_{kl}$ \cite{Cross2015}. Under pure unitary dynamics where the mediators return fully to their ground states ($|ij\rangle_\mathcal{A} \otimes |0,0\rangle_\mathcal{B} \rightarrow e^{i\phi_{ij}} |ij\rangle_\mathcal{A} \otimes |0,0\rangle_\mathcal{B}$), the values of $\beta_{ij,kl}$ remain purely real. In the presence of decoherence or residual mediator population, the phases $\beta_{ij,kl}$ become complex. We extract the net entangling phase via $\phi_{ZZ} = \textrm{Re}[\beta_{11, 00} - \beta_{10, 00} - \beta_{01, 00}]$, which maps to $\phi_{ZZ} = \phi_{11} + \phi_{00} - \phi_{10} - \phi_{01}$ for ideal unitary evolution.
By initializing the system in a complete qubit superposition state $|\psi(0)\rangle = \frac{1}{2}(|0, 0\rangle_\mathcal{A} + |1, 0\rangle_\mathcal{A} + |0, 1\rangle_\mathcal{A}+ |1, 1\rangle_\mathcal{A}) \otimes |0, 0\rangle_\mathcal{B}$, $\phi_{ZZ}$ can be extracted efficiently from a single master equation execution.

To match the simulations to the experimental data in Fig.~\ref{fig:cphase_mechanism}(d), we calibrate the $2\pi$-pulse amplitude on the target state across varying drive detunings and durations and calculate $\phi_{ZZ}$. Real experimental conditions include state preparation (SP) errors; notably, qubit mode $\mathcal{A}_1$ exhibits a high thermal population ($p_{\text{th}} = 0.23$) due to a poorly thermalized flux bias line. To ensure a good match with the data, we explicitly simulate the experimental measurement sequence [Fig.~\ref{fig:calibration}(e)] using the phases obtained from simulating purely unitary dynamics, alongside SP errors. Crucially, when the target entangling phase is exactly $\pi$, the extracted phase is invariant to SP errors, rendering it robust for calibrating a CZ gate. The resulting simulated phases [Fig.~\ref{fig:cphase_mechanism}(d)] show excellent agreement with experimental data across all pulse parameters without using any free fitting parameters.

\begin{figure}[H]
    \centering
    \includegraphics{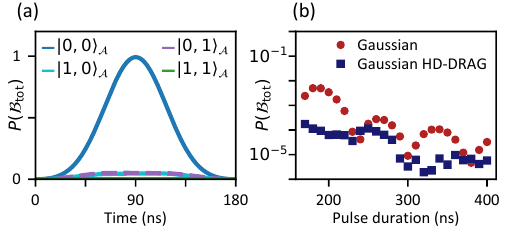}      
 \caption{Simulated total mediator population under unitary evolution. (a) Time evolution during a 180~ns Gaussian HD-DRAG pulse for different initial qubit states $|i, j\rangle_\mathcal{A}$. (b) Residual population after Gaussian (red) and Gaussian HD-DRAG (blue) pulses as a function of pulse duration.}
\label{fig:mediator_dynamics}
\end{figure}

To model the benchmarked gate fidelity in Sec.~\ref{sec:benchmarking}, we compute the average gate fidelity \cite{Nielsen2002}
\begin{align}
    \mathcal{F}(\mathcal{E}, U) = \frac{\sum_n \textrm{tr}(\hat{U} \hat{U}_n^\dagger \hat{U}^\dagger \mathcal{E}(\hat{U}_n)) + d^2}{d^2(d+1)},
\end{align}
where $\mathcal{E}$ is the quantum channel of our gate operation, $\hat{U}$ is the target unitary, $d=4$ and $\hat{U}_n$ are elements of the $d^2$-dimensional operator basis $\{ |i,j\rangle_\mathcal{A} \langle k, l|_\mathcal{A} \otimes |0,0\rangle_\mathcal{B} \langle 0, 0|_\mathcal{B} \; | \; i,j,k,l = 0, 1 \}$. For each pulse shape, parameters are optimized under ideal unitary dynamics by minimizing the infidelity $1 - \mathcal{F}$ using a SciPy-implemented BFGS optimizer \cite{SciPy2020}. Fig.~\ref{fig:mediator_dynamics}(a) shows the time evolution of a calibrated 180~ns HD-DRAG pulse conditioned on the initial qubit state $|i,j\rangle_\mathcal{A}$. The final state yields the residual total mediator population $P(\mathcal{B}_{\textrm{tot}}) = \langle \sum_\mu \hat{a}_{\mathcal{B}\mu}^\dagger \hat{a}_{\mathcal{B}\mu} \rangle$, plotted in Fig.~\ref{fig:mediator_dynamics}(b) for Gaussian and Gaussian HD-DRAG pulses across varying durations. The increase in residual population at shorter pulse durations corresponds to the coherent errors discussed in Sec.~\ref{sec:benchmarking}. Open-system master equation evaluations including the different collapse operators are then performed on the optimized controls to yield the average gate fidelity shown in Fig.~\ref{fig:cphase_benchmarking}(a) and the corresponding error budgets in Fig.~\ref{fig:cphase_benchmarking}(c, d).

\section{Single qubit gates \label{app:single_qubit gates}}
\begin{figure}[H]
    \centering
    \includegraphics{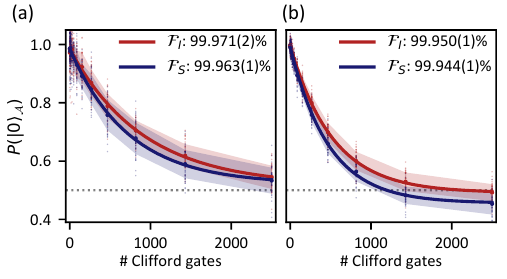}      
 \caption{Simultaneous randomized benchmarking (RB) for (a) Mode $\mathcal{A}_1$ of P-mon 1 and (b) Mode $\mathcal{A}_2$ of P-mon 2. Here, $\mathcal{F}_{I}$ and $\mathcal{F}_{S}$ denote the individual and simultaneous fidelities, respectively. The gray dashed line indicates a ground state population of 0.5}
\label{fig:simultanteous_RB}
\end{figure}

Single-qubit gates are implemented using 80 ns Gaussian DRAG pulses \cite{Motzoi2009, Werninghaus2021}, with the DRAG parameter calibrated to minimize phase errors. To evaluate performance, we perform simultaneous randomized benchmarking (RB) \cite{Gambetta2012}. The extracted individual ($\mathcal{F}_I$) and simultaneous ($\mathcal{F}_S$) fidelities are shown in Fig.~\ref{fig:simultanteous_RB}. For qubit mode $\mathcal{A}_1$, we find $\mathcal{F}_{I}=99.971(2)\%$ and $\mathcal{F}_{S}=99.963(1)\%$, while for qubit mode $\mathcal{A}_2$, the fidelities are $\mathcal{F}_{I}=99.950(1)\%$ and $\mathcal{F}_{S}=99.944(1)\%$. The comparable performance under both conditions indicates that the qubits can be addressed simultaneously with minimal degradation. We attribute the minor discrepancy between $\mathcal{F}_I$ and $\mathcal{F}_S$ to microwave crosstalk and the residual $ZZ$ interaction of $\zeta = 3.6(5)$~kHz. Notably, the simultaneous RB curve for mode $\mathcal{A}_2$ exhibits leakage, since it does not decay to the ideal ground-state population of 0.5 [dashed gray line in Fig.~\ref{fig:simultanteous_RB}]. Given the relatively long 80 ns pulse duration, this leakage likely originates from microwave crosstalk, as the $| 1 \rangle_{\mathcal{A}_2} \leftrightarrow | 2 \rangle_{\mathcal{A}_2}$ transition of mode $\mathcal{A}_2$ is close in frequency to the $| 0 \rangle_{\mathcal{A}_1}  \leftrightarrow | 1 \rangle_{\mathcal{A}_1}$ transition of mode $\mathcal{A}_1$ in this device (see Sec.~\ref{sec:device}).

\section{CZ gate calibration \label{app:calibration}}
\begin{figure}[H]
    \centering
    \includegraphics{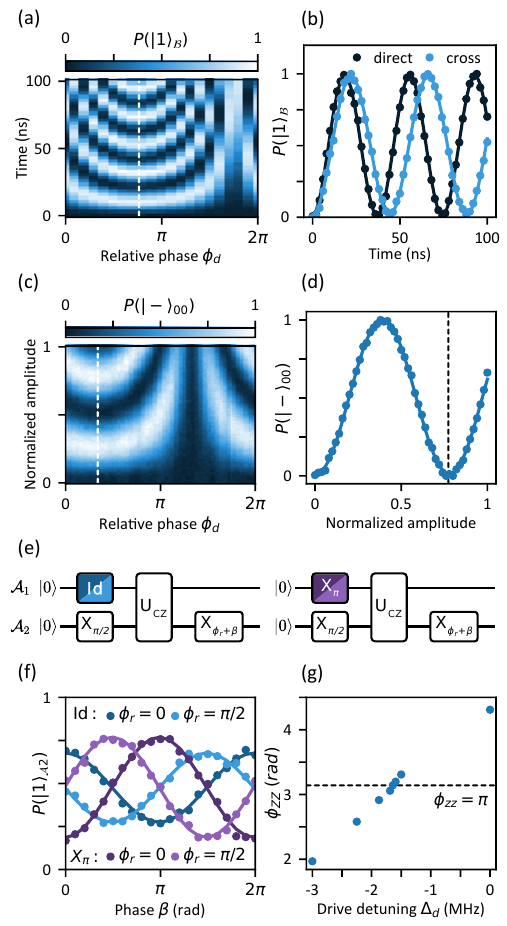}      
 \caption{(a) Time-domain Rabi oscillations of mode $\mathcal{B}_2$ under simultaneous drive (P-mon 1 \& 2) vs. relative drive phase $\phi_d$. (b) Comparison of mode $\mathcal{B}_2$ Rabi oscillations via direct (P-mon 2) vs. cross-drive (P-mon 1) line. (c) Amplitude Rabi oscillations of coupled mediator state $|-\rangle_{00}$ with crosstalk compensation vs. $\phi_d$; dashed line indicates the phase maximizing the Rabi rate. (d) Amplitude Rabi oscillations of state $|-\rangle_{00}$ identifying the $2\pi$ pulse amplitude (dashed line). (e) Gate sequences used to measure the acquired entangling phase from a unitary $U_{\text{CZ}}$. (f) Measured entangling phase as a function of the final phase $\beta$ for four distinct cases: applying either an identity ($\text{Id}$) or an $X_{\pi}$ gate, for final phase increments of $\phi_r = 0$ and $\pi/2$. (g) Entangling phase $\phi_{ZZ}$ vs. iteratively adjusted drive detuning $\Delta_d$, used to calibrate the specific detuning that implements a CZ gate (where $\phi_{\rm ZZ} = \pi$).}
\label{fig:calibration}
\end{figure}

We drive the CZ gate via the charge lines of P-mon 1 and P-mon 2. To exploit the symmetry of the target state (as discussed in Appendix~\ref{app:mediator_dynamics}), we need to calibrate the amplitude and relative phase of the two drives. Due to significant microwave crosstalk present in the device, it is crucial to pre-compensate the input signals accordingly. The microwave crosstalk matrix is defined as
\begin{align}
\boldsymbol M = 
\begin{pmatrix}
1 & a_{12}e^{i\phi_{12}} \\
a_{21}e^{i\phi_{21}} & 1
\end{pmatrix},
\end{align}
where $\phi_{12}$ and $\phi_{21}$ are the relative phases, and $a_{12}$ and $a_{21}$ are the amplitude ratios. To measure $\phi_{21}$, for example, we park mode $\mathcal{B}_2$ at its gate operation bias $\phi_{g2}$ and detune mode $\mathcal{B}_1$ by setting its corresponding flux bias to $\phi_1 = 0$. We then simultaneously drive Rabi oscillations on $\mathcal{B}_2$ via both the direct drive line (P-mon 2) and the cross-drive line (P-mon 1). By sweeping the relative phase of the two drives, we extract $\phi_{21}$ as the phase that maximizes the Rabi rate as shown in Fig.~\ref{fig:calibration}(a).
In principle, the amplitude ratio can also be extracted from this same measurement \cite{Haisheng2023}, however, we find it more robust to calibrate it separately.
This is achieved via Rabi oscillations using the direct and cross-drive lines shown in Fig.~\ref{fig:calibration}(b). Analogously, $\phi_{12}$ and $a_{12}$ are measured via mode $\mathcal{B}_1$. Given the crosstalk matrix $\boldsymbol M$, we adjust the drive signals $\boldsymbol \Omega_t = (\Omega_{t1}, \Omega_{t2})^T$ by precompensating our input drive signals via $\boldsymbol \Omega = (\Omega_1, \Omega_2)^T = \boldsymbol M^{-1} \boldsymbol \Omega_t$.

For the subsequent calibrations, we park both modes $\mathcal{B}_1$ and $\mathcal{B}_2$ at the gate operation bias $(\phi_{1g}, \phi_{2g})$. We first calibrate the relative phase $\phi_d$ of the precompensated drive signals to maximize the Rabi rate on the target transition $|00\rangle_\mathcal{B} \leftrightarrow |-\rangle_{00}$. Fig.~\ref{fig:calibration}(c) shows the population $P(|-\rangle_{00})$ versus drive amplitude and relative phase $\phi_d$. The maximum Rabi rate is indicated by a dashed line, and a full suppression of Rabi oscillations occurs when the phase is shifted by $\pi$, as expected from the dynamics described in Appendix~\ref{app:mediator_dynamics}. 

To calibrate the CZ gate, we must determine the drive amplitude and detuning $\Delta_d$ from the target transition required to generate the target entangling phase $\phi_{ZZ} = \pi$. First, we fix the drive detuning and calibrate the $2\pi$-pulse amplitude via Rabi oscillations on the target transition, as shown in Fig.~\ref{fig:calibration}(d). Next, we measure the acquired entangling phase $\phi_{ZZ}$ using the gate sequence presented in Fig.~\ref{fig:calibration}(e), with the corresponding measurement displayed in Fig.~\ref{fig:calibration}(f). Note that the population $P(|1\rangle_{\mathcal{A}2})$ of these measurements does not span the full 0 to 1 range because of a high thermal population ($p_{\rm th} = 0.23$) of mode $\mathcal{A}_1$. Furthermore, the variation in range between the different curves arises from readout crosstalk.
For each initial operation $\{\rm{Id}, X_{\pi}\}$ applied to mode $\mathcal{A}_1$, we increment the phase of the final gate $X_{\phi_r + \beta}$ by $\phi_r = 0, \pi/2$ and fit both oscillations simultaneously to achieve a more robust fit. The first sequence, which includes an initial $\rm{Id}$ gate on mode $\mathcal{A}_1$, yields the phase difference $\phi_{01} - \phi_{00}$. The second sequence, starting with an initial $X_\pi$ gate, yields the phase difference $\phi_{11} - \phi_{10}$. The entangling phase is then extracted via the relation $\phi_{ZZ}=\phi_{11} + \phi_{00} - \phi_{10} - \phi_{01}$. Based on the measured $\phi_{ZZ}$, the drive detuning is iteratively adjusted to converge on the target phase $\phi_{ZZ}=\pi$, as illustrated in Fig.~\ref{fig:calibration}(g). 
\noindent Once the entangling phase is set, the residual single-qubit phases are measured for both qubits using the first sequence shown in Fig.~\ref{fig:calibration}(e). These phases are then compensated via virtual $Z$ gates \cite{McKay2017} to finalize the CZ unitary.

For the Gaussian HD-DRAG pulse, we additionally calibrate the DRAG parameter. The drive signal consists of an in-phase component $\Omega_{\rm I}$ and a quadrature component $\Omega_{\rm Q}$ defined as:
\begin{align}
\Omega_{\rm I}(t) \propto [g(t) + c_1 \ddot{g}(t)], \;
\Omega_{\rm Q}(t) \propto c_2 [\dot{g}(t) + c_1 \dddot{g}(t)].
\end{align}
In the frequency domain, spectrum of $\Omega_{\rm IQ} = \Omega_I + i \cdot \Omega_Q$ takes the form:
\begin{equation}
\hat{\Omega}_{\rm IQ}(f) \propto [1 - (2\pi f)^2 c_1] \cdot [1 - (2\pi f) c_2] \cdot \hat{g}(f).
\end{equation}
By setting the coefficients $c_1 = 1/\delta_t^2$ and $c_2 = 1/\delta_t$, we introduce higher-order spectral suppression at the target detuning $\delta_t$. We initialize $\delta_t$ based on the measured frequencies of the closest non-target transitions $\omega_{10/01,+}$, calibrate the pulse amplitude and detuning as described previously, and finally refine the pulse parameters using a closed-loop optimizer \cite{Glaser2025}.

\newpage

%

\end{document}